\documentclass[letterpaper,twocolumn,10pt]{article}
\usepackage{usenix2019_v3}

\usepackage{tikz}
\usepackage{amsmath}
\usepackage{cite}
\usepackage{float} 
\usepackage{array}
\usepackage{textcomp}
\usepackage{stfloats}
\usepackage{url}
\usepackage{verbatim}
\usepackage{graphicx}
\usepackage{algorithmic}
\usepackage{algorithm}
\usepackage{caption}
\usepackage{subcaption}
\usepackage{multirow, makecell}
\usepackage{xspace}
\usepackage{graphicx}
\usepackage{filecontents}

\newcommand{\ie}{\emph{i.e.},\xspace}

\newcommand{\eg}{\emph{e.g.},\xspace}

\newcommand{\sysname}{{\sf AdaBridge}\xspace}

\newcommand{\parahead}[1]{\vspace*{0.4ex plus 0.15ex minus 0.15ex}\noindent %
  {\bfseries #1.}}

\renewcommand{\paragraph}[1]{\vspace{2pt plus 0pt minus 2pt}\noindent{\bfseries #1}}
\begin{document}
%-------------------------------------------------------------------------------

%don't want date printed
\date{}

\title{\Large \bf AdaBridge: Dynamic Data and Computation Reuse for Efficient Multi-task DNN Co-evolution in Edge Systems}

%for single author (just remove % characters)
% \author{
% {Lehao Wang}\\
% Northwestern Polytechnical University
% \and
% {Zhiwen Yu}\\
% Northwestern Polytechnical University and Harbin Engineering University
% \and
% {Sicong Liu}\\
% Northwestern Polytechnical University
% \and
% {Xiangrui Xu}\\
% Northwestern Polytechnical University
% \and
% {Chenshu Wu}\\
% The University of Hong Kong
% \and
% {Bin Guo}\\
% Northwestern Polytechnical University
\author{\rm Lehao Wang$^{1}$, Zhiwen Yu$^{1,2}$, Sicong Liu$^{1}$, Chenshu Wu$^{3}$, Xiangrui Xu$^{1}$, Bin Guo$^{1}$ \\ \\
\textsuperscript{1}Northwestern Polytechnical University, \textsuperscript{2}Harbin Engineering University,
\textsuperscript{3}The University of Hong Kong
}

% \affil[1]{Northwestern Polytechnical University}
% \affil[2]{Harbin Engineering University}
% \affil[3]{The University of Hong Kong}

% {2\quad the City University of Hong Kong, Hong Kong {\rm 999077}, China.}

% {\rm Second Name}\\
% Second Institution
% copy the following lines to add more authors
% \and
% {\rm Name}\\
%Name Institution
% } % end author

\maketitle

\begin{abstract}
Running multi-task DNNs on mobiles is an emerging trend for various applications like autonomous driving and mobile NLP. 
Mobile DNNs are often compressed to fit the limited resources and thus suffer from degraded accuracy and generalizability due to data drift. 
DNN evolution, \eg continuous learning and domain adaptation, has been demonstrated effective in overcoming these issues, mostly for single-task DNN, leaving multi-task DNN evolution an important yet open challenge. 
To fill up this gap, we propose \sysname, which exploits computational redundancies in multi-task DNNs as a unique opportunity for dynamic data and computation reuse, thereby improving training efficacy and resource efficiency among asynchronous multi-task co-evolution in edge systems. 
Experimental evaluation shows that \sysname achieves 11\% average accuracy gain upon individual evolution baselines. 

\end{abstract}

\section{Introduction}
\label{sec:lab} 
% Given the increasing proliferation of smart IoT devices, 

Deploying multi-task DNNs on diverse mobile devices enhances the efficient handling of various user interactions, resulting in a seamless and intelligent experience. For instance, smartphones or tablets can engage with a virtual assistant offering a range of services. However, due to limited memory and computation resources, mobile DNNs are typically compressed, impacting their generalization and making them susceptible to data drift~\cite{1}, leading to a drop in agnostic accuracy.
To address this, several approaches for DNN evolution have been explored, including continuous learning, online learning, and source-dependent/free domain adaptation with cloud or edge assistance. 
Edge-assisted continuous DNN evolution refines the mobile DNN with real-time sensor data on the mobile client, under the guidance of a global model at the edge server. 
This approach yields benefits such as increased accuracy, additional resources, and reduced latency.

However, either individual or multi-task DNN evolution methods struggle to balance evolution accuracy and efficiency in resource-constrained edge systems.
Applying them to \textit{multi-task} cases poses the following challenges:
\begin{itemize}
    \item 
    Each mobile client relying solely on local sensor data can impact evolution generalization, especially when the data doesn't satisfy the assumption of independent and identically distributed (IID) or is imbalanced. 
    Despite local DNN, our key insight is that the sensor data from each mobile client has the potential to enhance the evolution accuracy of DNNs on other clients. 
    However, uploading all client sensor data introduces transmission overhead and complicates edge-evolution training. 
    Moreover, resampling suitable sensor data across multiple mobile models, especially in the absence of DNNs or data distribution of other tasks, remains a significant challenge.

    \item It is challenging to achieve efficient co-evolution of multiple asynchronous tasks with limited edge memory and computing resources. 
    We note computational similarities and redundancies in retraining models from different clients, serving as both \textit{burdens} and \textit{opportunities} for data and computation reuse. 
    The dynamic nature of asynchronous multi-task evolution requests makes these redundancies and opportunities constantly changing. 
    Also, balancing the urgency and resource demands of multiple tasks complicates achieving average high accuracy, low-cost, and flexible co-evolution of them.
\end{itemize}

These challenges motivate the design of \sysname.
As far as we know, it is the first work to adaptively bridge dynamic data and computation reuse across algorithm and system co-designed levels.
It can simultaneously improve training efficacy and resource efficiency among asynchronous multi-task co-evolution in edge systems.

\section{AdaBridge Design}
\label{sec:design}

\begin{figure*}[t]
% \vspace{-1cm}
    \centering
    \includegraphics[width=.93\textwidth]{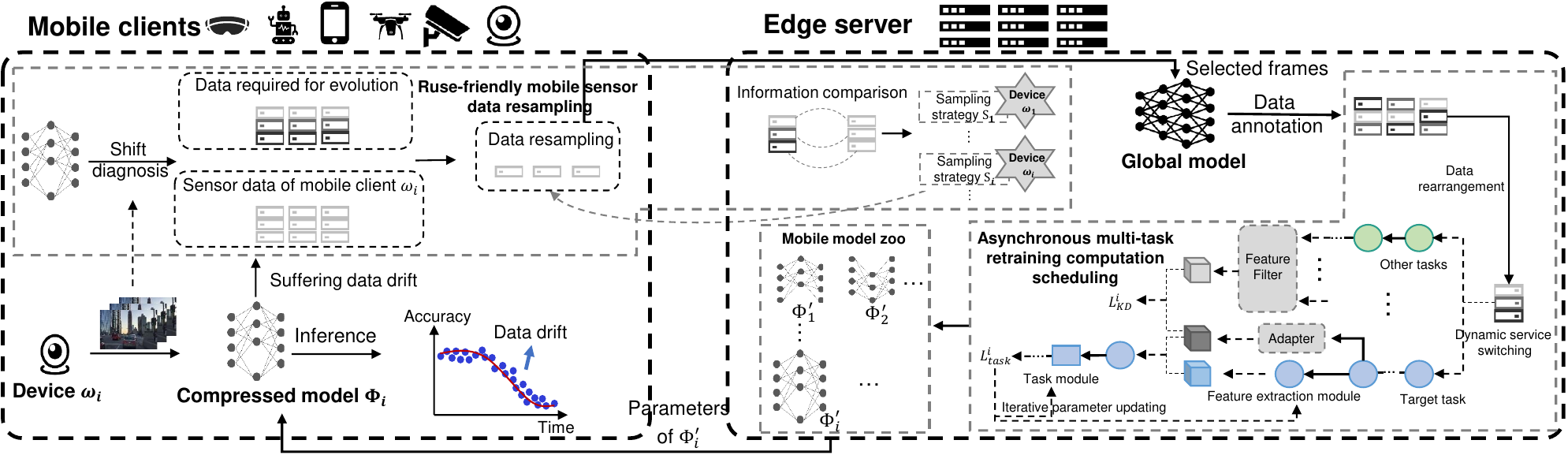}
    % \vspace{-4cm}
    \caption{Illustration of \sysname's system workflow.}
    \vspace{-2mm}
    \label{fig:overview}
\end{figure*}

As shown in Figure~\ref{fig:overview}, 
Core components of \sysname include a \textit{reuse-friendly mobile sensor data resampling} module and an \textit{asynchronous multi-task retraining computation scheduling} module. 
We present the detailed module design as follows.

\parahead{Reuse-friendly mobile sensor data resampling}
\label{sec:device}
It is non-trivial for mobile ends to dynamically resample locally sensed data useful for the evolution of not only themselves but also other client DNNs. 
The reasons are two-fold:
\textit{i)} prior research faces difficulties in rapidly evaluating the \textit{explicit data contribution} for accuracy gain of local DNNs with new, unlabeled sensor data \textit{before actual training execution}. 
This problem is more complex when predicting \textit{implicit} data contribution for accuracy gains for DNN evolution on other clients that are locally inaccessible. 
Achieving accurate and real-time profiling is challenging.
\textit{ii)} in the absence of distribution of sensor data from other clients, determining the \textit{implicit complementarity} between local data and other clients' data is intricate.
Therefore, mobile clients first conduct data shift type analysis based on runtime accuracy profiling for multi-task DNNs, and then generate the $\mathrm{X} \to \mathrm{Y}$ mapping function of other DNNs for data distribution comparison and resampling.

 \begin{figure}[t]
  \hfill
  \subfloat {\includegraphics[width=0.2\textwidth]{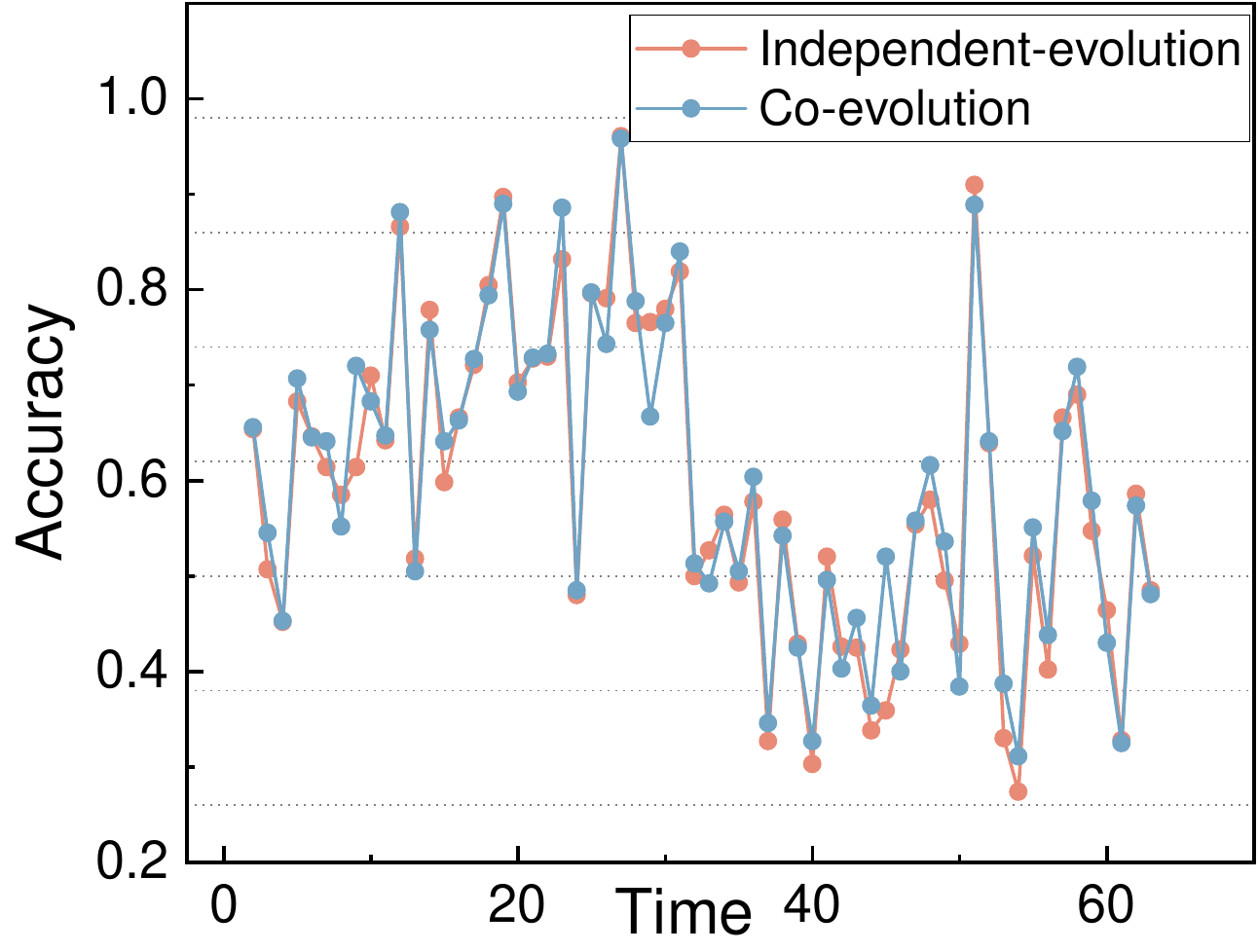}}
  \hfill
  \subfloat 
  {\includegraphics[width=0.2\textwidth]{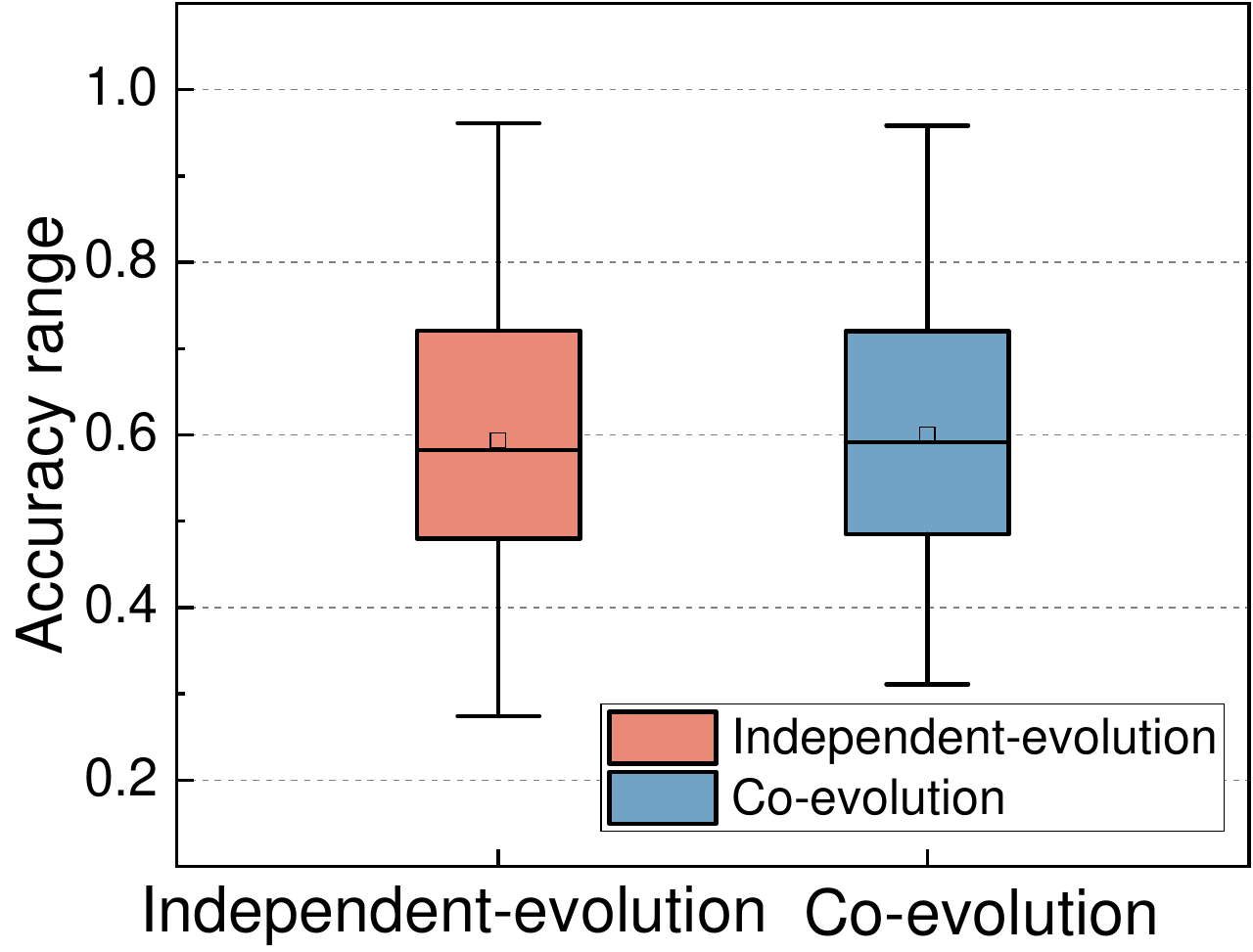}}
  \hfill
    \hspace{0.4cm}
  \caption{Comparison of accuracy fluctuation.}
\vspace{-5mm}
  \label{fig:ex}
 \end{figure}
 
\parahead{Asynchronous multi-task retraining computation scheduling}
\label{sec:server}
Upon receiving data from mobile devices, the edge server faces challenges in improving data reuse for multitask training, computation scheduling, and resource allocation.
% Multi-task learning encounters difficulties in flexibly integrating reused results and separating them for diverse tasks to deploy to clients effectively. 
% Additionally, the increased I/O reads and writes resulting from similar searches in computation reuse during training contribute to high latency.
% Thus, 
We propose the following design:
\textbf{i) Plug-and-Play Asynchronous Multi-task Computation Reusing}: To achieve data reuse and accommodate flexible model retraining fusion and separation, we introduce an adapter-based multi-task learning framework. 
It asynchronously reuses learned features from different subtasks and incorporates dynamic switching and incremental parameter updating to optimize the limited representation capacity of compressed mobile DNNs.
\textbf{ii) Memory I/O Cost-aware Underlying Computation Reuse}: To further enhance I/O efficiency in computation reuse, \sysname reorganizes data before training. It reuses nearby features with high similarity, reducing the frequency of reads and writes to high-level memory and decreasing computational loads.

\section{Experiment}
\label{sec:experiment}
We have conducted a preliminary experiment in this section. 

\parahead{Experimental settings}
We adopt two vehicles as mobile clients, \ie mobile unmanned cars (A), and unmanned aerial vehicles (UAV) (B), and deploy compressed Faster-RCNN with ResNet50 on them, respectively. 
Two NVIDIA GeForce RTX3080 GPUs with 10GB memory serve as the edge server.
And they capture live video while riding on open roads.

\parahead{Deployment and results}
We compare the evolution performance of independent evolution with local data and \sysname's co-evolution with data reuse.
Figure.~\ref{fig:ex} shows the results.
\sysname performs better when facing data drift with 11\% improvement in lowest accuracy. 
This is because the co-evolved DNN has better generalization ability. 
The co-evolved DNN outputted by \sysname is stable in accuracy, contributing to the high flexibility and separability. 

\section{Conclusion}
\label{sec:conclusion}
This paper presents \sysname, an end-to-end multi-task DNN co-evolution system designed for resource-constrained edge systems. 
By establishing a flexible adaptive reuse bridge between multiple tasks, the system efficiently handles the continuous evolution requests of asynchronous mobile DNNs with limited resources, aiming to overcome data drift.
\bibliographystyle{plain}
% \bibliography{sample.bib}

\end{document}